\documentclass[12pt]{article}
\usepackage{epsfig,axodraw}
\usepackage{epsfig}
\begin{document}
%%%%%%%%%%%%%%%%%%%%%%%%%%%%%%%%%%%%%%%%%%%%%%%%%%%%%%%%%%%%%%%%%%%%%%%%%%%
\thispagestyle{empty}
\begin{flushright}
{MRI-P-030701}\\
{IC/2003/53}\\
hep-ph/0307117 \\
\end{flushright}
\begin{center}
{\LARGE\bf Bhabha Scattering with Radiated Gravitons at Linear Colliders}
\\
\vspace*{0.4in}
{\sl Sukanta Dutta$^a$, 
\footnote{On vacation leave, Department of Physics, S.G.T.B. Khalsa College, 
University of Delhi, Delhi, India. }, 
Partha Konar$^b$, Biswarup Mukhopadhyaya$^b$,
{\rm and} Sreerup Raychaudhuri$^c$} \\
\vspace*{0.3in}
\begin{tabular}{l}
$^a$ The Abdus Salam International Centre for Theoretical Physics, Trieste, 
Italy. \\
~E-mail: {\sf Sukanta.Dutta@cern.ch} \\ 
$^b$Harish-Chandra Research Institute, Chhatnag Road, Allahabad 211019, 
India. \\
~E-mail: {\sf konar@mri.ernet.in, biswarup@mri.ernet.in} \\ 
$^c$Department of Physics, Indian Institute of Technology, Kanpur 208016,
India. \\
~E-mail: {\sf sreerup@iitk.ac.in} \\
\end{tabular}

\vspace*{1.5cm}
{\Large\sc Abstract}
\end{center}
\begin{quotation}
\noindent
We study the process $e^+  e^- \to e^+ e^-  \not{\!\!E}$ at a high-energy
$e^+ e^-$ collider, where the missing 
energy arises from the radiation of Kaluza-Klein gravitons in a model with
large extra dimensions. It is shown that at a high-energy linear collider,
this process can not only confirm the signature of such theories but can also
sometimes be comparable in effectiveness to the commonly discussed channel
 $e^+ e^- \to  \gamma  \not{\!\!E}$, especially for a large number of extra 
dimensions and with polarized beams. We also suggest some ways of 
distinguishing the signals of a graviton tower from other types of new 
physics signals by combining data on our suggested channel with those on 
the photon-graviton channel.
\end{quotation}
\vskip 1 cm


\vfill
%%%%%%%%%%%%%%%%%%%%%%%%%%%%%%%%%%%%%%%%%%%%%%%%%%%%%%%%%%%%%%%%%%%%%%%%%%%
\newpage
\section{Introduction}

It is a matter of common belief in the high-energy physics community
that there must be physics beyond the Standard Model (SM) around the TeV
scale. This belief is founded in the
fact that some of the parameters of the SM can be made stable against
quantum corrections only by invoking new physics at the TeV scale.
With this in mind, an enormous international effort is being poured
into the construction of the 14-TeV LHC at CERN, which is expected to run by
2008 and come up with signals for new physics. However, as the
LHC is a hadronic machine, it is unlikely to yield signals which are
unequivocal as to the underlying theory. It is therefore widely 
deemed necessary to build a high-energy $e^+ e^-$ collider to confirm the
nature and detailed properties of this new physics.

A high-energy $e^+ e^-$ collider operating in the range 500~GeV to a few
TeV must be of the linac type to avoid large synchrotron radiation
losses. Such a machine would require a very high luminosity in order to be
competitive with other accelerators. This is, in fact, an essential feature 
of the
design\cite{LC} of proposed linear colliders. The relatively clean environment
of an $e^+ e^-$ collider allows the complete reconstruction of individual 
events, making it possible to carry out precision tests of new physics. 
Moreover, one has
the facility of polarizing the beams, which can play a significant role
in background elimination. It is also possible to run such a machine in the
$e^- e^-$, $e\gamma$ and $\gamma\gamma$ modes. In this paper we consider the
possibility of using a high energy  $e^+ e^-$ collider to detect signatures
of low-scale quantum gravity in a model with large extra dimensions. In
particular, we focus on Bhabha scattering with radiated gravitons.

One of the most exciting theoretical developments of recent years has
been the idea that there could be one or more extra spatial dimensions
and the observable Universe could be confined to a four-dimensional
hypersurface in a higher-dimensional `bulk' spacetime\cite{Akama}. 
Such ideas, which can be  motivated by superstring theories, 
give rise to elegant solutions\cite{ADD} 
to the well-known gauge hierarchy problem of high
energy physics, which is just the instability against quantum
corrections mentioned above. What is even more interesting, perhaps,
is the suggestion\cite{GRW,HLZ} 
that there could be observable signals of quantum
gravity at current and future accelerator experiments, and this
possibility has spawned a vast and increasing body of work over the
past five years. This relatively new set of ideas, commonly dubbed 
`Brane World Phenomenology', bases itself on two main principles: the
concept of hidden compact dimensions and the string-theoretic idea of 
$D_p$-branes. The simplest brane-world scenario is the so-called
Arkani-Hamed---Dimopoulos---Dvali (ADD) model\cite{ADD}, in which there 
are $d$ extra spatial dimensions, compactified on a $d$-torus of radius
$R_c$ each way. Together with the four canonical Minkowski
dimensions, this constitutes the `bulk' spacetime. In this scenario the 
radius $R_c$
of the extra dimensions can be as large as a quarter of a 
millimeter\cite{LCP}.
However, the SM fields are confined to a four-dimensional slice of
spacetime, with thickness not more than $10^{-17}$~cm, which is called
the `brane'. If the ADD model is embedded in a string-theoretic
framework, the `brane' is, in fact a $D_3$-brane, i.e. a 3+1
dimensional hypersurface on which the ends of open strings are
confined\cite{Polchinksi}. 
However, it is not absolutely essential to embed the model in
a string theory, and the word `brane' or `wall' is then used simply to
denote the hypersurface (or thin slice) where the SM fields are confined.
A crucial feature of this model is that gravity, which is a property of
spacetime itself, is free to propagate in the bulk. As a result
\begin{itemize}
\item Planck's constant in the bulk $M_S$, identified with the `string 
scale', and fixed by the string tension $\alpha'$, is related to Planck's 
constant on the brane $M_P$ ($\simeq~1.2\times 10^{19}$~GeV) by
\begin{equation}
(M_S)^{2+d} = (4\pi)^{d/2} ~\Gamma(d/2) ~M_P^2 ~(R_c)^{-d}
\end{equation}

\noindent
which means that for $R_c \sim 0.2$~mm, it is possible to have
$M_S$ as low as a TeV for $d \geq 2$. (The normalization of
reference \cite{HLZ} has been adopted here). This solves the gauge 
hierarchy problem 
simply by bringing down the scale of new physics (i.e. strong gravity in this case) 
to about a TeV and thereby providing a natural cut-off to the SM, 
since the string scale $M_S$ now controls graviton-induced processes on the brane.

\item There are a huge number of massive Kaluza-Klein excitations of the
(bulk) graviton field on the brane, with masses $m_n = n/R_c$, 
and these collectively
produce gravitational excitations of electroweak strength, which may be
observable at current experiments and those planned in the near 
future\cite{Kubyshin}.
\end{itemize}

It is only fair to mention that a major drawback of the ADD model is
that it creates a new hierarchy between the `string scale' $M_s \sim 1$~TeV 
and the size of the extra dimensions $R_c^{-1} \sim
1$~$\mu$eV. In fact, the huge size of the extra dimensions (compared
to the Planck length) is not stable under quantum corrections, which
tend to shrink it down until $M_s \sim R_c^{-1} \sim M_P \sim
10^{19}$~GeV, at which stage the original hierarchy problem is
reinstated. Nevertheless, there are several variants of
the ADD model which address this problem in various ways, and some of
these ideas may not be far from the truth.  From a phenomenological
point of view, it is, therefore, reasonable to postpone addressing the
stability issue, and proceed to study the minimal ADD model and its
consequences for experiment.

The experimental consequences of ADD gravity have been mainly studied
in the context of of ($a$) real graviton emission and ($b$) virtual 
graviton exchange. In the former case
the final state gravitons in the ADD model are `invisible', escaping the
detector because of their feeble individual interactions ($\sim M_P^{-1}$)
with matter. The final state, involving missing energy due to gravitons, 
will be built up by making an
incoherent sum over the tower of graviton modes. In the case of virtual
gravitons, the final state is built up by making a coherent sum. In
either case, it may be shown\cite{GRW,HLZ} that, after summing, the 
Planck mass $M_P$ cancels
out of the cross-section, leaving an interaction of near-electroweak strength.

At an $e^+ e^-$ collider, the most-frequently 
discussed\cite{GRW,MPP,Wilson} signal 
for ADD gravitons is the process $e^+ e^- \to \gamma G_n$,
where $G_n$ is the $n$th Kaluza-Klein excitation of the graviton field.
This process leads to single photon events with missing energy and momentum,
and is expected to be among the earliest signals of ADD gravity at an $e^+ e^-$
collider. In addition, the process $e^+ e^- \to  \mu^+ \mu^- G_n$ has also
been studied\cite{Eboli}. In this work, we have studied the process
$$
e^+ e^- \to  e^+ e^- G_n
$$
which is simply Bhabha scattering with a radiated graviton. The final state
would contain an electron-positron pair with substantial missing energy.
The reasons for taking up a study of this process are as follows.
\begin{itemize}
\item At an $e^+ e^-$ collider, final states with an $e^+ e^-$ pair will be
one of the first things to be analyzed, since this is needed for beam 
calibration\cite{LC}.
Among such states, it should be a simple matter to select events with large missing
energy as well.
\item Unlike  $e^+ e^- \to  \mu^+ \mu^- G_n$, this process does not suffer from
large $s$-channel suppression, especially if one considers final state 
electrons (positrons) in relatively forward (backward) directions.
\item Taken in conjunction with the process $e^+ e^- \to \gamma G_n$, 
this process
could help to distinguish signals of the ADD scenario from other physics models
beyond the SM. This is elaborated in the subsequent discussions.
\item In some ways, as we shall see, this process has signatures more distinct from
other new physics effects than $e^+ e^- \to \gamma G_n$.
\end{itemize}
In the following section we describe the salient features of our calculation
of the signal process. Backgrounds and some strategies for their elimination
are discussed in section 3. Section 4 contains the results of our numerical
analysis. In section 5 we discuss how this process can be used to pinpoint
the model if signatures of the suggested type are indeed observed. We summarize
and conclude in the final section.

%%%%%%%%%%%%%%%%%%%%%%%%%%%%%%%%%%%%%%%%%%%%%%%%%%%%%%%%%%%%%%%%%%%%%%%%%%%
\section{The Process $e^+ e^- \to  e^+ e^- G_n$}

As gravity couples to the energy-momentum tensor, each component of the tower 
of ADD gravitons couples to all SM fields, as well as to each SM interaction 
vertex. The relevant Feynman rules can be found, for example, in 
Refs.\cite{GRW,HLZ}. Using these, it can be shown that the process
$e^+ e^- \to  e^+ e^- G_n$ is driven, at tree-level, by 28 Feynman diagrams. 
Some 
representative diagrams are shown in Fig.~1, which are obtained by `dressing'
an $s$-channel process $e^+ e^- \to  e^+ e^-$ with graviton radiation. 
Fig.~1($a$) represents a 
graviton emission from one of the external electron legs; there will be 
4 such diagrams with the graviton emitted from {\it each} leg in turn. 
Fig.~1($b$) represents a graviton emission from the gauge boson propagator. 
Fig.~1($c$) represents a graviton emission from one vertex; there will
be another such with graviton emission from the other vertex. Thus Fig.~1 
encompasses 14 diagrams, 7 each with $Z$ and photon exchanges. In addition, 
there will be
another set of 14 diagrams with the $\gamma,Z$ exchange in the $t$-channel,
making 28 diagrams in all. This latter set is absent in the case of
$e^+ e^- \to  \mu^+ \mu^- G_n$.

%===============================Fig. 1====================================
{
\unitlength=1.3 pt
\SetScale{1.25}
\SetWidth{0.5}      % line    size control
\scriptsize    %  letter  size control
% 1
\begin{picture}(90,90)(0,0)
\ArrowLine(0,20)(30,40)
\ArrowLine(30,40)(0,60)
\Photon(30,40)(60,40){4}{3}
\ArrowLine(60,40)(90,20)
\ArrowLine(90,60)(60,40)
\Photon(15,50)(45,70){3}{5}
\Photon(45,70)(15,50){-3}{5}
\Text(0,15)[l]{\large $e^-$}
\Text(0,70)[l]{\large $e^+$}
\Text(45,25)[c]{\large $\gamma ,Z$}
\Text(100,15)[r]{\large $e^+$}
\Text(100,70)[r]{\large $e^-$}
\Text(59,75)[r]{\large $G$}
\Text(45,8)[c]{\large $(a)$}
\end{picture} \
{} \qquad\allowbreak
% 2
\begin{picture}(90,90)(0,0)
\ArrowLine(0,20)(30,40)
\ArrowLine(30,40)(0,60)
\Photon(30,40)(60,40){4}{3}
\ArrowLine(60,40)(90,20)
\ArrowLine(90,60)(60,40)
\Photon(45,40)(45,70){3}{5}
\Photon(45,70)(45,40){-3}{5}
\Text(0,15)[l]{\large $e^-$}
\Text(0,70)[l]{\large $e^+$}
\Text(45,25)[c]{\large $\gamma ,Z$}
\Text(100,15)[r]{\large $e^+$}
\Text(100,70)[r]{\large $e^-$}
\Text(45,80)[c]{\large $G$}
\Text(45,8)[c]{\large $(b)$}
\end{picture} \
{} \qquad\allowbreak
% 3
\begin{picture}(90,90)(0,0)
\ArrowLine(0,20)(30,40)
\ArrowLine(30,40)(0,60)
\Photon(30,40)(60,40){4}{3}
\ArrowLine(60,40)(90,20)
\ArrowLine(90,60)(60,40)
\Photon(60,40)(90,40){3}{5}
\Photon(90,40)(60,40){-3}{5}
\Text(0,15)[l]{\large $e^-$}
\Text(0,70)[l]{\large $e^+$}
\Text(45,25)[c]{\large $\gamma ,Z$}
\Text(100,15)[r]{\large $e^+$}
\Text(100,70)[r]{\large $e^-$}
\Text(105,40)[r]{\large $G$}
\Text(45,8)[c]{\large $(c)$}
\end{picture}}
\begin{center}
Figure 1.
{\it Representative Feynman diagrams for the process $e^+ e^- \to  e^+ e^- G_n$.}
\end{center}
%=======================================================================

Considering the process
$$
e^-(k_1,\lambda_1) ~e^+(k_2,\lambda_2) \to  G_n(p_1) ~e^+(p_2) ~e^-(p_3)
$$
with a single Kaluza-Klein mode $G_n(p_1)$ in the final state, we obtain
the unpolarized cross-section 
\begin{equation}
\sigma(m_n) = \frac{1}{2s} \int 
\frac{d^3\vec{p_1}}{(2\pi)^3 ~2E_1}  
~\frac{d^3\vec{p_2}}{(2\pi)^3 ~2E_2}
~\frac{d^3\vec{p_3}}{(2\pi)^3 ~2E_3} 
~\delta^4(k_1 + k_2 - p_1 - p_2 - p_3)
~~\frac{1}{4}\sum_{\lambda_1, \lambda_2} |M_n(\lambda_1, \lambda_2)|^2
\label{cross}
\end{equation}
where the final state helicities are summed over. This cross-section
is a function of the mass $m_n$ of the graviton excitation.

The squared matrix element $|M_n(\lambda_1, \lambda_2)|^2$ for the process 
can be written as the sum
\begin{equation}
|M_n(\lambda_1, \lambda_2)|^2 = 
\left|\sum_{i = 1}^{28} M_n^{(i)}(\lambda_1, \lambda_2)\right|^2
\end{equation}
where $|M_n^{(i)}(\lambda_1, \lambda_2)|^2$ represents the squared helicity 
amplitude arising from the $i$th diagram. 
In our calculation, we make use of the helicity amplitude technique to 
write down amplitudes for all these 28 diagrams and evaluate the individual
terms in the above sum using the subroutine HELAS\cite{Hagiwara}. 

Recalling that the observed final state, namely, $e^+ e^-$ plus missing energy,
consists of an 
incoherent sum over the tower of graviton modes, we get
\begin{eqnarray}
\sigma(e^+ e^- \to e^+ e^- + \not{\!\!E}) 
& = & \sum_n \sigma(m_n) \nonumber \\
& \simeq & \int_0^{\sqrt{s}} dm ~\rho(m) ~\sigma(m)
\end{eqnarray}
approximating the discrete (but closely-spaced) tower of states by a continuum.
The density of states is given by\cite{HLZ}
\begin{equation}
\rho(m) = \frac{2~R_c^d ~m^{d-1}}{(4\pi)^{d/2} ~\Gamma(d/2)} \ ,
\end{equation}
and the integration is cut off at the kinematic limit $\sqrt{s}$.

As the calculation, even using HELAS, is long and 
cumbersome, some checks on the numerical results are called for.
The most useful check is provided by the Ward identities arising from 
general coordinate invariance, which constitute
an essential feature of any theory involving gravity. 
We can write the amplitude for the emission of any  graviton in the form
\begin{equation}
M_n(\lambda_1, \lambda_2) 
= T^{\mu\nu}(\lambda_1, \lambda_2) ~\epsilon^{(n)*}_{\mu\nu}(p_1)
\end{equation}
where $\epsilon^{(n)}_{\mu\nu}(p_1)$ is the polarization tensor for the
$n$th (massive) graviton mode. The tensor 
$T^{\mu\nu}(\lambda_1, \lambda_2)$ is {\it the same for every mode}, including 
the massless mode $\epsilon^{(0)}_{\mu\nu}(p_1)$, which is the usual graviton
of four-dimensional Einstein gravity. This must now satisfy the Ward
identities
\begin{equation}
p_1^\mu T_{\mu\nu}(\lambda_1,\lambda_2)
= p_1^\nu T_{\mu\nu}(\lambda_1,\lambda_2) = 0 \ ,
\end{equation}
where we note that
\begin{eqnarray}
T^{\mu\nu}(\lambda_1, \lambda_2) = \sum_{i = 1}^{28}
T_i^{\mu\nu}(\lambda_1, \lambda_2)
\end{eqnarray}
with $i$ indicating the $i$th diagram, as above. 
The consistency check
therefore
requires a perfect cancellation, for each choice of
$\lambda_1$ and $\lambda_2$, between all such terms\footnote{In fact the 
$s$ and $t$-channel diagrams 
form two independent gauge-invariant sets, so that the actual cancellation 
takes place between 14 diagrams at a time.}, which is highly sensitive to 
errors in signs and factors.
We have also checked that if only the $s$-channel results are taken
our numerical results are in close agreement with those of Ref.\cite{Eboli},
and this, of course, implies that the $t$-channel contributions can be
easily checked using crossing symmetry.

In our numerical analysis of this problem, we have also considered diagrams 
mediated by gravitons in place of $\gamma,Z$. Such diagrams, in fact, are 
of the same order in perturbation theory after summation over the 
graviton propagators, but it turns out that the
contributions are very small --- typically less than 1\% for 
experimentally-allowed values of $M_S$ --- due to the smallness 
of the effective graviton coupling after summation.

It is also worth mentioning that contributions due to graviscalars 
in the ADD model can also be neglected. It can be shown\cite{HLZ} 
that diagrams
with graviscalars coupling to external fermion legs and vertices
undergo large cancellations and the final contribution is proportional
to the electron mass. This leaves only the diagrams with graviscalars
emitted from the massive $Z$-propagator, which is again suppressed at high 
energies by a factor of $M_Z^2/s$  compared to the gravitensor 
contributions, the latter being jacked up by derivative couplings.

\section{Background Elimination}

The actual signal for the process considered in the previous section consists
of a hard electron and a hard positron, with substantial missing energy. 
The SM backgrounds arise from all processes of the form $e^+ e^- 
\to e^+ e^- \nu \bar{\nu}$, where the neutrinos can be of any flavor. 
At the lowest order, there are 96 Feynman diagrams which 
give rise to this final state. Some of the principal sub-processes are 
listed below.
\begin{eqnarray}
e^+ e^- & \to & e^\pm W^\mp \nu_e^{\!\!\!\!\!\!(-)} 
          \to e^+ e^- \nu_e \bar{\nu}_e \nonumber \\
e^+ e^- & \to & e^+ e^- Z \to  e^+ e^- (\nu \bar{\nu}) \nonumber \\
e^+ e^- & \to & \nu_e \bar{\nu}_e Z \to  \nu_e \bar{\nu}_e (e^+ e^-) \nonumber \\
e^+ e^- & \to & W^+ W^- \to (e^+ \nu_e) (e^- \bar{\nu}_e) \nonumber \\
e^+ e^- & \to & ZZ \to (e^+ e^-) (\nu \bar{\nu}) \nonumber \\
e^+ e^- & \to & \gamma^*Z \to (e^+ e^-) (\nu \bar{\nu}) 
\end{eqnarray}
Of course, for every process with real $W/Z$, there will also be many 
diagrams with off-shell particles.
Taken together, all these diagrams constitute a formidable background
to the suggested signal. A judiciously chosen set of kinematic 
cuts, however,
enable us to reduce these backgrounds enormously. For example, diagrams with
the neutrino pair (missing energy) arising from a real $Z$ boson can be easily 
removed
by putting a cut on the $\nu\bar{\nu}$ (missing) invariant mass. 
A similar cut on the $e^+ e^-$ pairs would seem obvious but it is not 
advisable, since it would 
remove a large part of the {\it signal} arising from radiative return to 
the $Z$-pole
through graviton emission. This, however, is not a serious problem, as the
background contribution due to $e^+ e^- \to  \nu_e \bar{\nu}_e Z \to  
\nu_e \bar{\nu}_e (e^+ e^-)$ is not very large. On the other hand, the
background due to the  process $e^+ e^- \to  e^\pm W^\mp 
~\nu_e^{\!\!\!\!\!\!(-)}$ is more problematic, as most cuts  tending to reduce
this also affect the signal adversely.

In principle, an additional source of  background can be the 
`two-photon' process $e^+ e^- \to e^+ e^- e^+ e^-$ 
where one electron-positron pair escapes detection by being emitted close 
to the beam pipe. Here one has to remember that such forward electrons, 
being rather hard (with energy on the order of $100$~GeV or above), should be 
detectable in the end caps of the electromagnetic calorimeter, although their 
energy resolution will be rather poor\cite{LC}. 
Thus it should be possible to impose 
a veto on hard electrons up to within a few degrees of the beam pipe, 
whereby the above 
background can be virtually eliminated. Events with the forward electrons even 
closer to the beam pipe\cite{Chen} are removed via a cut on missing $p_T$. 

Taking all these considerations into account, we focus on
an $e^+ e^-$ collider operating at a center-of-mass energy of 
500~GeV (1~TeV).
The initial set of kinematic cuts used in our  analysis are as follows:
\begin{itemize}
\item The final state electron (positron) should be at least $10^0$ away from the 
beam pipe. This tames the collinear singularities arising from $t$-channel
photon exchange. At the same time, it also ensures that any background
effects from beamstrahlung are mostly eliminated\cite{Godbole}.
\item The electron (positron) should have a transverse momentum 
$p_T^e > 10$~GeV.
\item We demand a missing transverse momentum $p_T^{\rm miss} > 15 (25)$~GeV
for $\sqrt{s} =$ 500~GeV (1~TeV). As mentioned above, this also helps in reducing
the two-photon background.
\item The tracks due to electron and positron must be well-separated, with
$\Delta R > 0.2$, where $R = \sqrt{\Delta\eta^2 + \Delta\phi^2}$, in terms
of the pseudorapidity $\eta$ and the azimuthal angle $\phi$.
\item The opening angle between the electron and positron tracks is required
to be limited by
$5^0 < \theta_{e^+e^-} < 175^0$. This ensures elimination of possible cosmic ray 
backgrounds, and also ensures  sufficient missing energy.
\item The missing invariant mass $M_{\rm miss}$ should satisfy the cut
$|M_{\rm miss} - M_Z| > 10$~GeV. This important cut eliminates many SM 
backgrounds in which two final state neutrinos arise from a $Z$-decay. 
\end{itemize}

%-------------------------------------------------------------------
\begin{figure}[h]
\setcounter{figure}{1}
\centerline{
\epsfxsize= 9.0 cm\epsfysize=8.0cm
                     \epsfbox{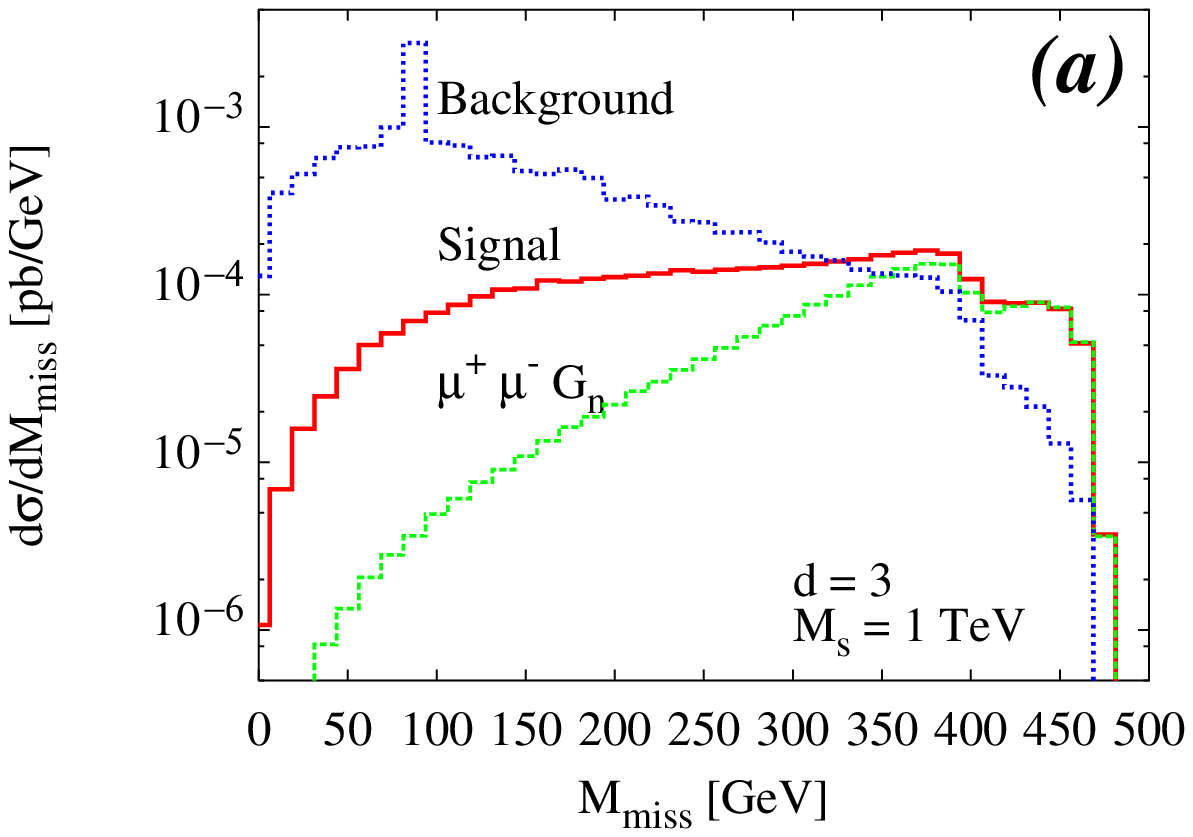}
        \hspace*{-0.4cm}
\epsfxsize=9.0 cm\epsfysize=8.0cm
                     \epsfbox{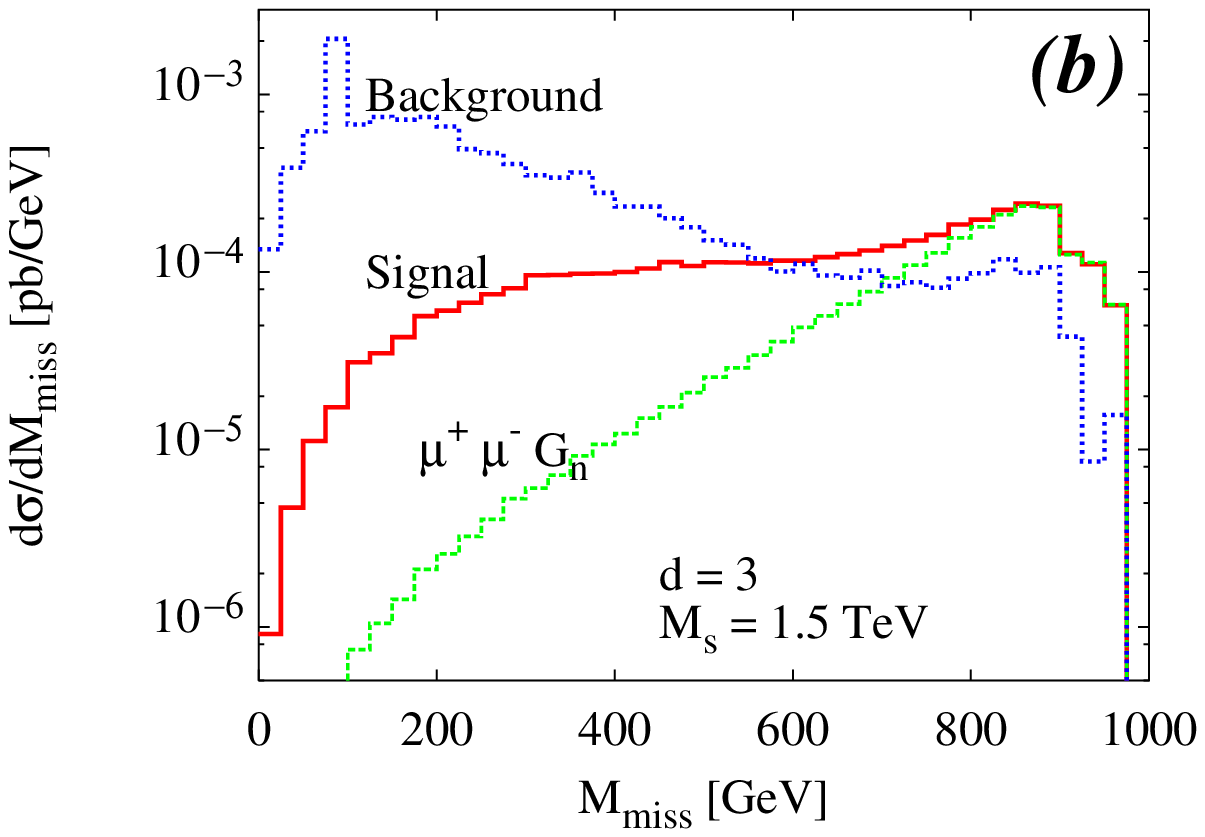}
}
\caption{{\footnotesize\it 
Distribution in missing invariant mass $M_{\rm miss}$ 
for $\sqrt{s} =$ ($a$)~500~GeV and ($b$)~1~TeV. In addition to total 
signal and background cross-sections, we also show the signal due to
$s$-channel diagrams only, which is identical with the cross-section 
for $e^+e^- \to \mu+\mu^-G_n$. 
}}
\label{fig:invm_dist}
\end{figure}
%-------------------------------------------------------------------

Incorporating all these cuts into a Monte Carlo event generator, 
we obtain the distribution in missing invariant mass shown 
in Figure~2. It is clear from the figure that the missing invariant mass
of the signal is much harder than that of the background. This is easy to
understand if we realize that the missing mass is a measure of the energy
of the emitted graviton. Now it is well-known that the higher the energy,
the more strongly the graviton is coupled to matter and consequently the higher
the production cross-section. Moreover, a higher energy leads to a higher
density of states, and this further enhances the cross-section. 
In Figure 2, we have also shown the value of the signal when only $s$-channel
diagrams are considered, which would be the case if we were to study processes 
like $e^+ e^- \to \mu^+ \mu^- G_n$ or $e^+ e^- \to \tau^+ \tau^- G_n$. 
Comparison with the signal of interest makes it 
obvious that there is a substantial contribution to the Bhabha
scattering signal from the $t$-channel, which makes it more viable as a
signal of ADD gravity than the other two. In fact, the emission of a 
high-energy
graviton (corresponding to large missing invariant mass) reduces the energy
of the underlying Bhabha scattering process, causing a large enhancement
due to the $Z$-pole. This is the reason why the signal for large $M_{\rm miss}$
is completely dominated by the $s$-channel contribution. At the same time,
when the graviton is relatively soft, the $s$-channel diagrams are strongly
suppressed, but the $t$-channel contributions are not; as a result the
signal is dominated by the $t$-channel contribution for low $M_{\rm miss}$.
We can therefore
obtain a clear separation of signal from background by imposing a cut
$$
M_{\rm miss} > 350 ~(450) ~{\rm GeV~~for}~\sqrt{s} = 500 ~{\rm GeV}~( 1 ~{\rm T
eV})
$$
These choices are optimal, after taking into account the fact that the signal
falls rapidly with increasing $M_s$ and $d$.

An important strategy for background reduction is the use of the beam 
polarization facility at a linear collider. If the electron (positron) beam
has a right (left) polarization efficiency ${\cal P}_e$ (${\cal P}_p$), the 
cross-section formula corresponding to equation \ref{cross} is obtained by 
the replacement
\begin{eqnarray}
\sum_{\lambda_1, \lambda_2} |M_n(\lambda_1, \lambda_2)|^2 
& \longrightarrow &
~~~(1 + {\cal P}_e)(1 - {\cal P}_p) |M_n(+,+)|^2
+ (1 + {\cal P}_e)(1 + {\cal P}_p) |M_n(+,-)|^2
\nonumber \\
&& +~(1 - {\cal P}_e)(1 - {\cal P}_p) |M_n(-,+)|^2
+ (1 - {\cal P}_e)(1 + {\cal P}_p) |M_n(-,-)|^2
\nonumber \\
\end{eqnarray}
Typical values for the polarization efficiencies\cite{LC} 
used in this analysis are 
${\cal P}_e = 0.8$ and ${\cal P}_p = 0.6$, which tend to favor the first
two terms on the right side of the above expression. The use of polarized
beams leads to a drastic reduction in the SM background, chiefly because
it causes suppression of the $W$-induced diagrams. The numerical effects are
presented in the next section. 

\section{Results and Discussions}

After numerical evaluation of signal and background, we find it convenient
to present our results, using the significance $S/\sqrt{B}$, where $S (B) = 
{\cal L} \sigma_{S(B)}$, for given values of $d$ and $M_S$,
Here $\sigma_{S(B)}$ denotes the
cross-section for the signal (background), while ${\cal L}$ is the integrated
luminosity. Our numerical results are presented assuming 
${\cal L} = 500$~fb$^{-1}$; however, it is a simple matter to scale the 
significance values for different values of luminosity.

In Figure 3($a$) we show curves showing the variation of the
significance with the string scale $M_S$, for $d = 2, 3, 4 ,5$ and $6$ 
respectively, for $\sqrt{s} = 500$~GeV. Figure 3($b$) shows a similar
plot for $\sqrt{s} = 1$~TeV. In both cases, we have considered unpolarized
electron and positron beams.

%-------------------------------------------------------------------
\begin{figure}[h]
\centerline{
\epsfxsize= 9.0 cm\epsfysize=8.0cm
                     \epsfbox{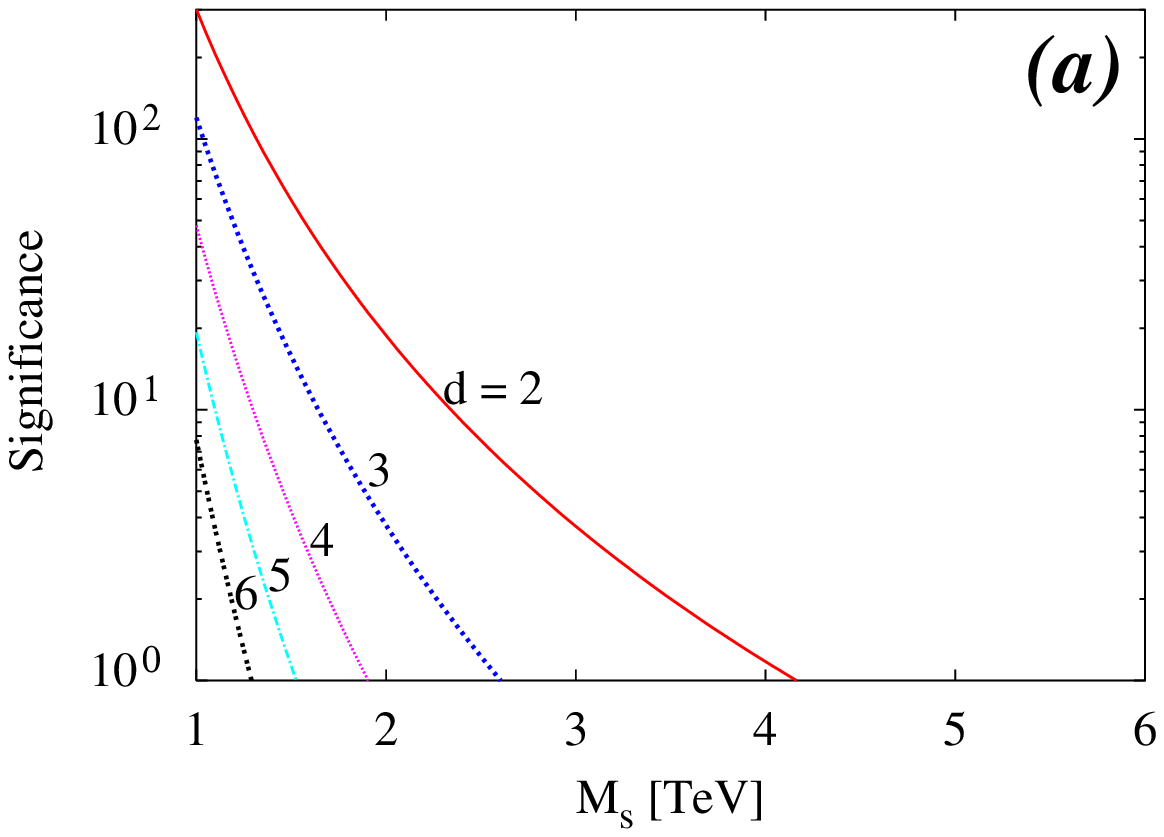}
        \hspace*{-0.7cm}
\epsfxsize=9.0 cm\epsfysize=8.0cm
                     \epsfbox{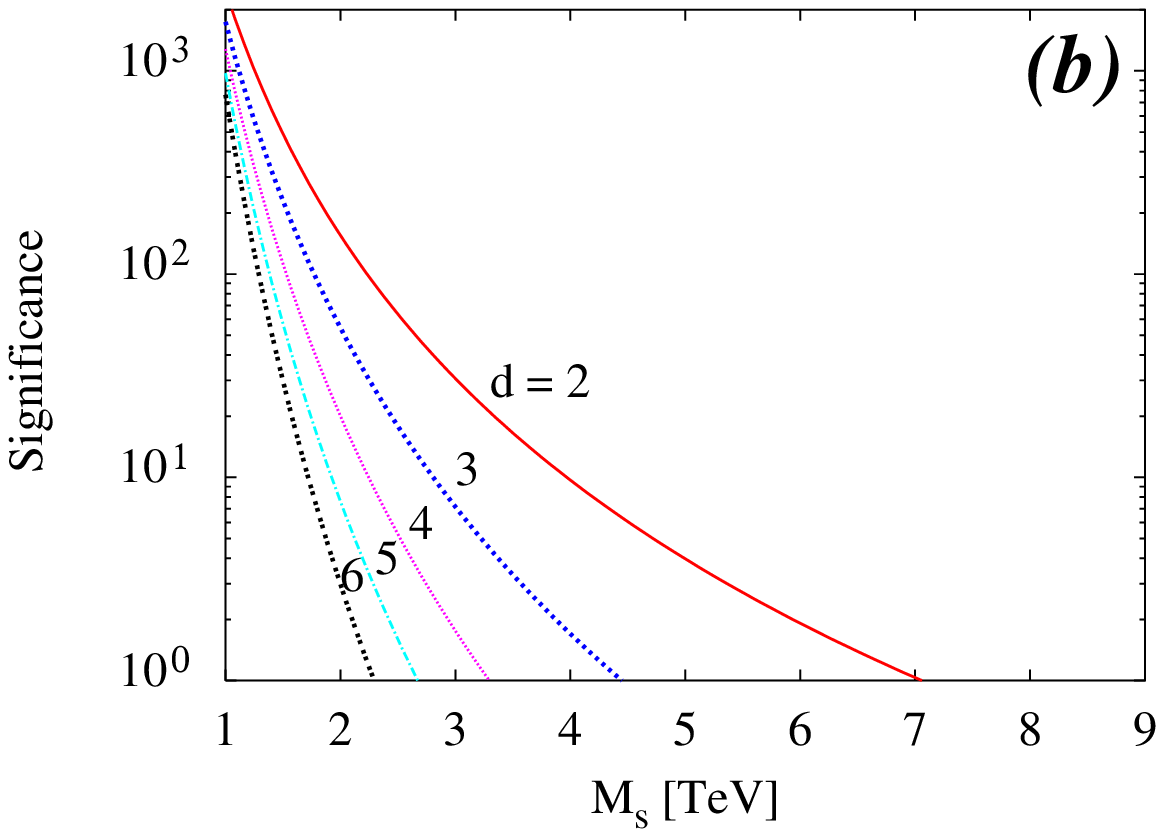}
}

\caption{{\footnotesize\it 
Illustrating the variation of significance $S/\sqrt{B}$ with 
$M_S$, considering unpolarized
beams for ($a$) $\sqrt{s} = 500$~GeV, and ($b$)  $\sqrt{s} = 1$~TeV. 
}}
\label{fig:signif1}
\end{figure}
%-------------------------------------------------------------------

As has been already mentioned, the use of beam polarization improves 
changes of detecting the signal quite dramatically. This is illustrated in 
Figure 4, which is similar to Figure 3, except that the polarization 
efficiencies have been taken to be ${\cal P}_e = 0.8$ and ${\cal P}_p = 0.6$ 
for the electron and  positron respectively.

%-------------------------------------------------------------------
\begin{figure}[h]
\centerline{
\epsfxsize= 9.0 cm\epsfysize=8.0cm
                     \epsfbox{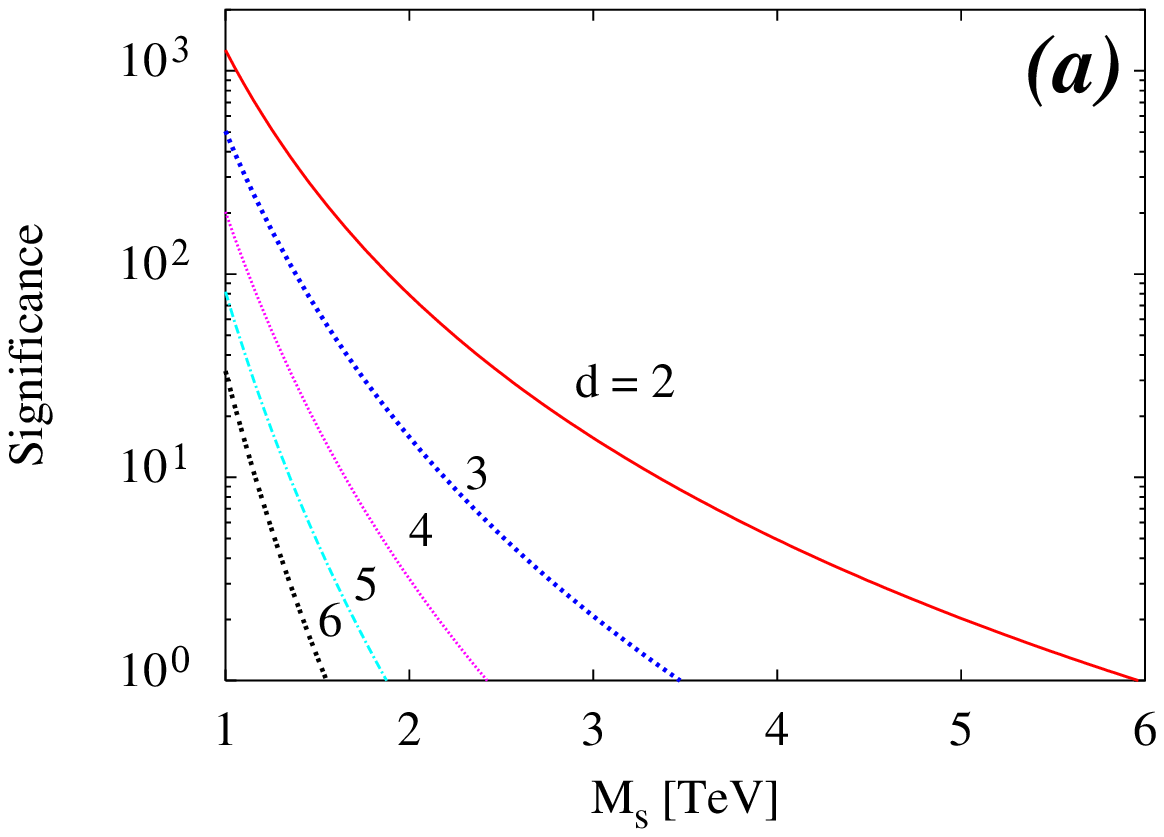}
        \hspace*{-0.7cm}
\epsfxsize=9.0 cm\epsfysize=8.0cm
                     \epsfbox{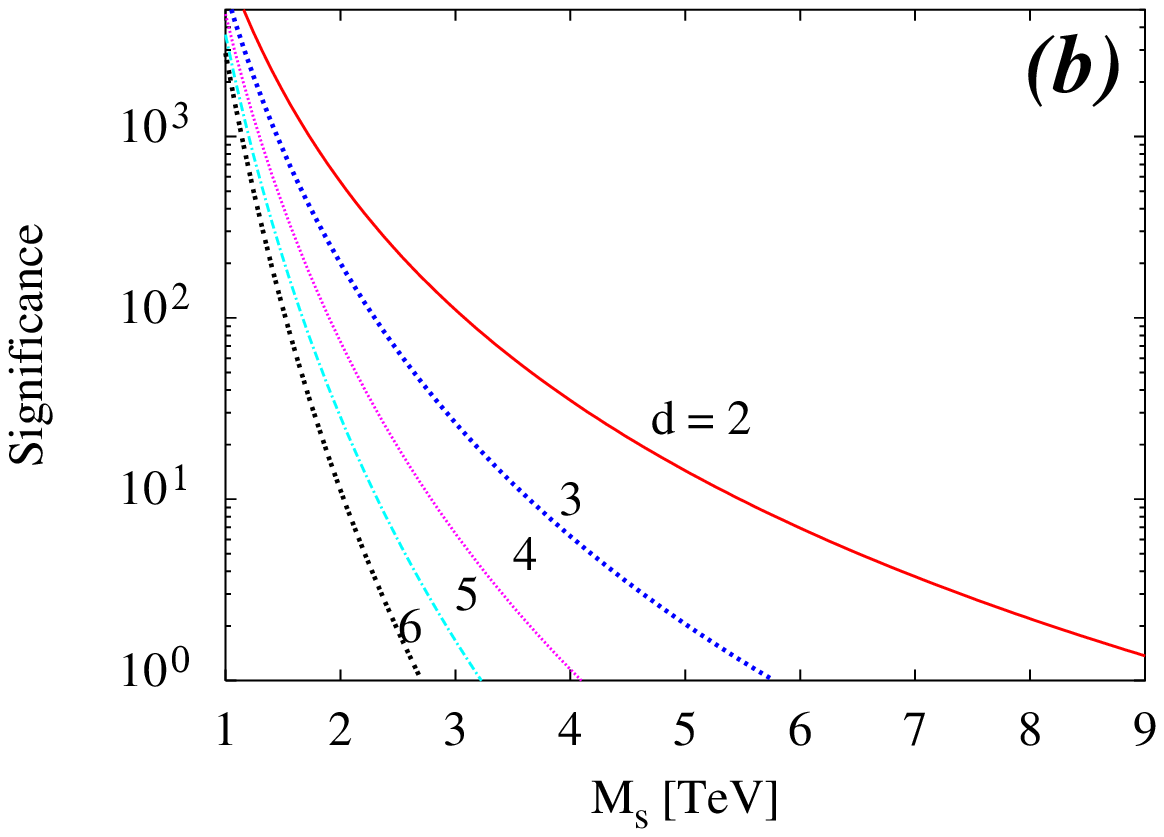}
}

\caption{{\footnotesize\it 
Illustrating the variation of significance $S/\sqrt{B}$ with
$M_S$, considering polarized
beams for ($a$) $\sqrt{s} = 500$~GeV, and ($b$)  $\sqrt{s} = 1$~TeV.
We take ${\cal P}_e = 0.8$ and ${\cal P}_p = 0.6$.
}}
\label{fig:signif2}
\end{figure}
%-------------------------------------------------------------------

From the above graphs, it is straightforward to  find out the maximum value
of $M_s$ that can be probed at the linear collider for any given value
of $d$, both for polarized and unpolarized beams. In Table 1, we
present the values of $M_s$ that one can probe  at 99\% confidence level.

%-------------------------------------------------------------------
\vspace*{-0.35in}
\begin{center}
$$
\begin{array}{|c|r|c|c|c|c|c|c|}
\hline
\sqrt{s}    
           & d =               & 2    & 3    & 4    & 5    & 6    \\ 
\hline\hline
0.5 
           & {\rm polarized}   & 4.53 & 2.79 & 2.01 & 1.60 & 1.35 \\
           & {\rm unpolarized} & 3.16 & 2.09 & 1.59 & 1.30 & 1.12 \\
\hline\hline
1.0
           & {\rm polarized}   & 7.40 & 4.63 & 3.41 & 2.76 & 2.36 \\
           & {\rm unpolarized} & 5.36 & 3.57 & 2.74 & 2.28 & 2.36 \\
\hline
\end{array}
$$
\end{center}
\noindent
Table 1.
{\footnotesize\it 
99\% C.L. discovery limits on the string scale $M_S$ for an 
ADD scenario 
using (graviton) radiative Bhabha scattering, for different numbers of extra 
dimensions. An integrated luminosity of $500~fb^{-1}$ has been assumed. All
numbers are in TeV.}
%-------------------------------------------------------------------
\vskip 10pt

If one compares these with the corresponding reach of the 
process\cite{GRW,MPP,Wilson}
$e^+ e^- \to \gamma + \not{\!\!E}$, one will notice that search limits
are approximately of the same order. However,
the comparison should not be made
too literally, because of several reasons. First, the polarization
efficiencies
in our case, while matching with those of \cite{Wilson}, are different from
those
of \cite{GRW}, while the center-of-mass energy at which reference
\cite{Wilson} has calculated
the effects is slightly different from ours. In addition, one needs to
match the event selection criteria more carefully for a full-fledged
comparison.
Finally, it should be borne in mind that in our analysis we have
adopted a normalization of the string scale $M_s$, which, though widely used,
is not necessarily uniform in the literature. However, in spite of such
non-uniformities, Figures 3 and 4, together with Table 1,
can  perhaps be taken as faithful indications
of the fact that the predictions on radiative Bhabha scattering are comparable
to those on the photon-graviton channel, so far as the limits of the probe
on the string scale in ADD models are concerned.

\section{Pinning Down the Model}

A few comments are in order on how to distinguish graviton signals involving 
missing energy from similar ones arising from other kinds of new physics. For example,
let us consider the well-known process $e^+ e^- \to \gamma G_n$, leading to a
single-photon-plus-missing-energy signal. Such signals can be obtained in
several other models. A model with an extra $Z'$ boson could lead to a process
$e^+ e^- \to \gamma Z'$ with the $Z'$ decaying to neutrinos. Similarly, in
supersymmetric models, a process like $e^+ e^- \to \gamma \widetilde{\chi}^0_1
\widetilde{\chi}^0_1$ (where $\widetilde{\chi}^0_1$ is the lightest neutralino)
would also lead to a single photon signal. Although the $Z'$ signal can be
differentiated by looking for a resonant peak in the missing invariant mass,
the supersymmetric signal is more difficult to disentangle.

The situation for the process considered in this paper is somewhat more
encouraging. It is true that both types of new physics considered 
in the preceding paragraph
can produce the same signal. Processes of the form $e^+ e^- \to e^+ e^- Z'$,
with the $Z'$ decaying to neutrinos are one possibility. In supersymmetry, too, 
it is possible to pair-produce either sleptons or charginos, followed by decays
to electrons (positrons) and invisible particles. The $Z'$ signal can again
be differentiated by a peak in the missing invariant mass. The supersymmetric
signal, in this case, always arises from production of a pair of real sparticles,
which should emerge in opposite hemispheres and, if light enough, will be highly
boosted. As a result, the observed electron and positron should lie, most of
the time, in opposite hemispheres. For the graviton signal, however, there
is no such correlation; in fact the distribution in the opening angle between 
$e^+$ and $e^-$ turns out to be practically flat. This distinction
clearly does not work very well when the produced sparticles are heavy.
%-------------------------------------------------------------------
%\vspace*{1.7cm}
\begin{figure}[h]
\centerline{
\epsfxsize= 9.0 cm\epsfysize=8.0cm
                     \epsfbox{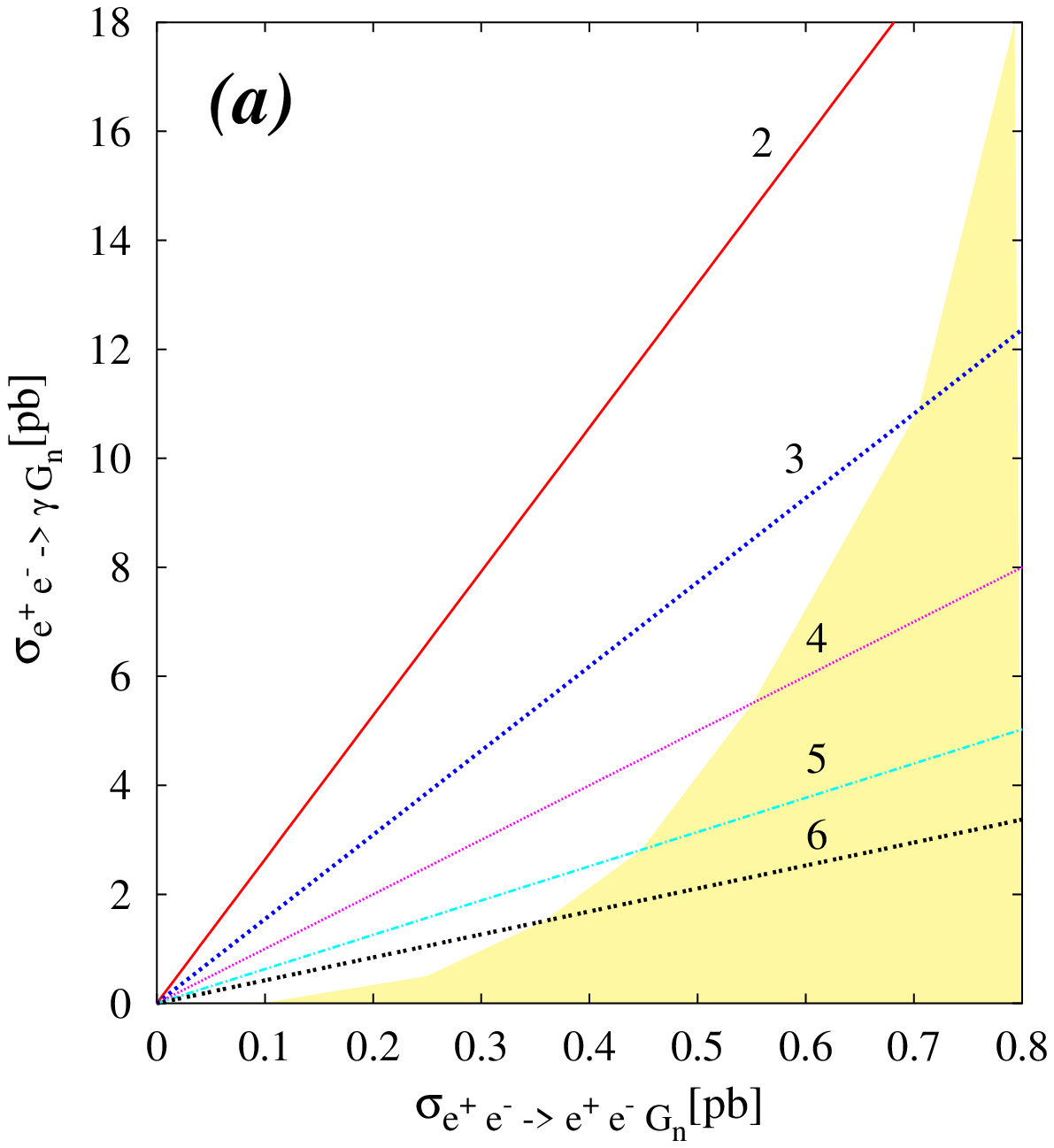}
        \hspace*{-0.7cm}
\epsfxsize=9.0 cm\epsfysize=8.0cm
                     \epsfbox{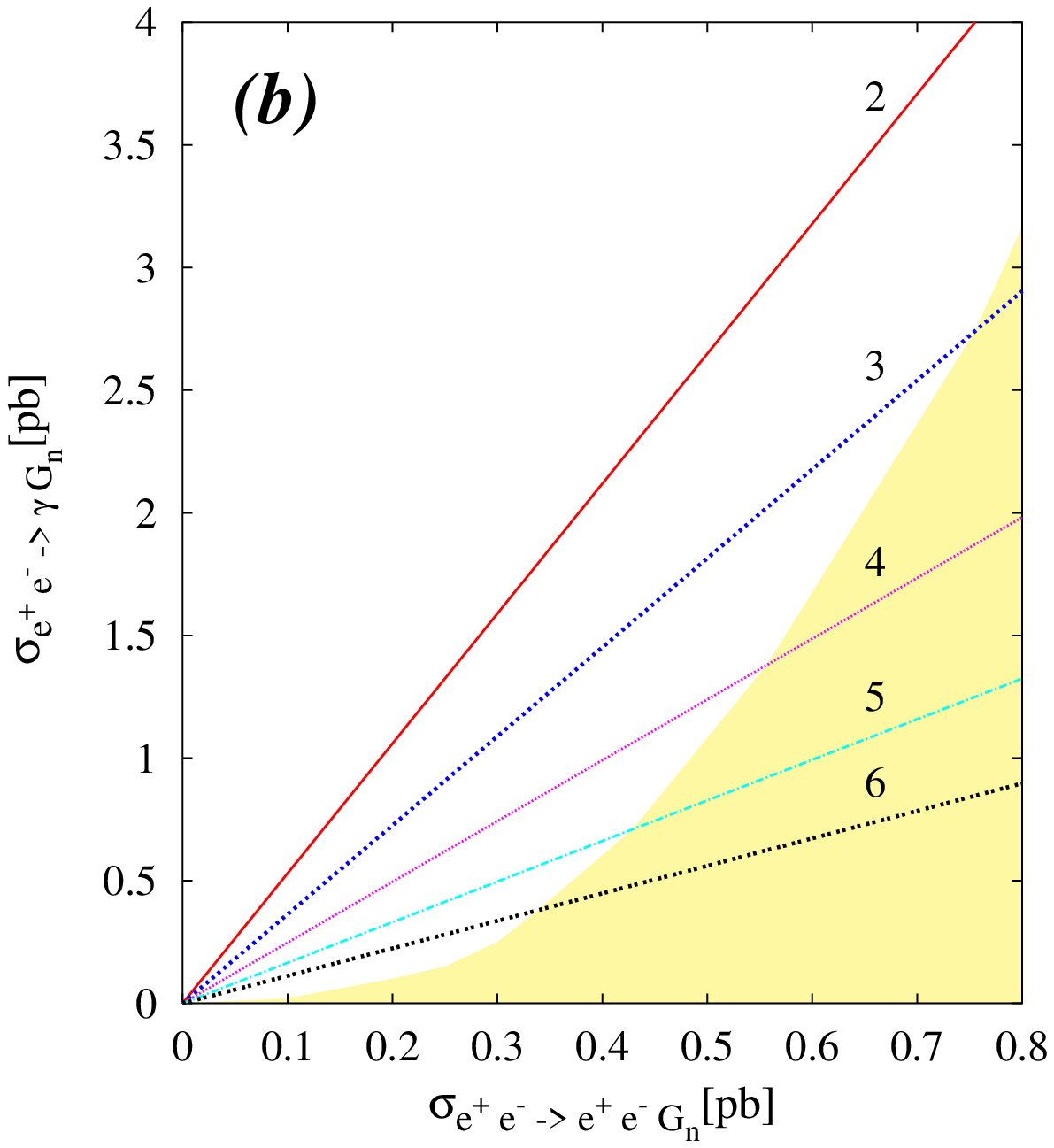}
}
\caption{{\footnotesize\it 
Correlation plot showing the polarized cross-sections for the
processes $e^+ e^- \to \gamma G_n$ and $e^+ e^- \to e^+ e^- G_n$
for different values of $d$. Each curve is generated by varying $M_S$
for ($a$) $\sqrt{s} = 500$~GeV, and ($b$)  $\sqrt{s} = 1$~TeV.
Shading indicates the region where the theoretical calculation is
unreliable.
}}
\label{fig:compare}
\end{figure}
%-------------------------------------------------------------------

A more elegant way of distinguishing these signals from those other 
forms of new
physics is to compare the cross-sections arising from {\it both} the processes
$e^+ e^- \to \gamma G_n$ and $e^+ e^- \to e^+ e^- G_n$, both of which are
determined by the two parameters $d$ and $M_S$, and hence, will have some
correlation. In Figure 5, we have plotted the cross-sections 
$\sigma(\gamma G_n)$ versus $\sigma( e^+ e^- G_n)$ for different values
of $d$. In view of the discussion in the previous section,
we would like to note that here the rates for $e^+ e^- \to \gamma G_n$
have been computed with the same normalization of the string scale as that 
used for the Bhabha scattering process. Each curve corresponds to variation 
of $M_S$ over the range allowed by current experimental constraints. 
Once experimental data 
are available from a high energy $e^+ e^-$ machine, we should be able to
pinpoint a small region (ideally a point) on the graph(s) in Figure 5.
The position of this would immediately direct us to the value of
$d$; comparison of the cross-sections would now yield a measurement
of the value of $M_S$. The identification of the model parameters should be
unambiguous since the curves for different values of $d$ do not intersect 
except at the origin. In any case, the area in the vicinity of the origin is 
limited by the search limits indicated in Table 1 since it corresponds to 
very high values of $M_S$. The shaded region corresponds to $M_S = \sqrt{s}$
beyond which the theoretical calculations are unreliable.
It should, of course be remembered that further information on 
the fundamental parameters of theory can be extracted from various kinematic 
distributions which are sensitive to these parameters.

%%%%%%%%%%%%%%%%%%%%%%%%%%%%%%%%%%%%%%%%%%%%%%%%%%%%%%%%%%%%%%%%%%%%%%%%%%%
\section{Summary and Conclusions}

In the paper we have considered the process $e^+ e^- \to e^+ e^- G_n$,
which is essentially Bhabha scattering with radiated gravitons in the ADD 
model. After identifying suitable event selection criteria, we find that
this process can act as an effective probe of large extra dimensions
at a high-energy $e^{+} e^{-}$ collider, especially with a center-of-mass
energy of order 1~TeV and with polarized beams. The string scale $M_S$ that 
can be probed in this channel is found to be comparable to that which 
is accessible through the alternative process $e^+ e^- \to \gamma G_n$. 
It is also shown that a study of the (graviton) radiative Bhabha scattering 
process may provide some further handle on the essential characteristics of 
ADD-like theories. And finally, taking a cue from the wisdom that, in coming 
to any conclusion on new physics possibilities, it is always advantageous to 
have more than one type of data, we have demonstrated how our predictions 
can be combined with those on the photon-graviton channel to obtain rather 
trustworthy revelations on models with large extra dimensions.

%%%%%%%%%%%%%%%%%%%%%%%%%%%%%%%%%%%%%%%%%%%%%%%%%%%%%%%%%%%%%%%%%%%%%%%%%%%
\bigskip
\centerline{\bf Acknowledgments}
{\footnotesize
The authors acknowledge useful discussions with D.Choudhury, R.M.Godbole,
U.Mahanta, and S.K.Rai. This work was initiated
as part of the activity of the Indian Linear Collider Working Group
(Project No. SP/S2/K-01/2000-II of the Department of Science and Technology, 
Government of India). SD thanks the SERC, Department of Science and Technology, Government of India for partial support. The work of BM was
partially supported by the Board of Research in Nuclear Sciences (BRNS),
Government of India. SR thanks the Harish-Chandra Research
Institute for hospitality while this paper was being written.
}

%%%%%%%%%%%%%%%%%%%%%%%%%%%%%%%%%%%%%%%%%%%%%%%%%%%%%%%%%%%%%%%%%%%%%%%%%%%

%%%%%%%%%%%%%%%%%%%%%%%%%%%%%%%%%%%%%%%%%%%%%%%%%%%%%%%%%%%%%%%%%%%%%%%%%%%
\end{document}